\documentclass[10pt,a4paper]{article}

\usepackage{amsmath,amssymb,amsthm}
\usepackage{url}
\usepackage{placeins}
\usepackage{subfigure}
\usepackage{multirow}
\usepackage{epsfig}
\usepackage{chicago}
\usepackage{rotating}

\newcommand{\Pbb}{\ensuremath{\mathbb{P}}}

\newcommand{\E}{\ensuremath{\mathbb{E}}}

\newcommand{\F}{\ensuremath{\mathcal{F}}{}\!}
\newcommand{\G}{\ensuremath{\mathcal{G}}{}\!}

\newcommand{\xVA}{\ensuremath{{\rm {xVA}}}}
\newcommand{\pd}{\ensuremath{{\rm {PD}}}}
\newcommand{\lgd}{\ensuremath{{\textsc{Lgd}}}}

\newcommand{\probSpace}{\ensuremath{(\Omega,\F,\ \Pbb)}}

\newcommand{\CVA}{\ensuremath{{\text{CVA}}}}

\newcommand{\I}[1]{\ensuremath{ I_{#1}  }}

\newcommand\be{$$}
\newcommand\ee{$$}
\newcommand\ben{\begin{equation}}
\newcommand\een{\end{equation}}
\newcommand\bea{\begin{eqnarray*}}
\newcommand\eea{\end{eqnarray*}}
\newcommand\bean{\begin{eqnarray}}
\newcommand\eean{\end{eqnarray}}

\begin{document}
\author{ Chris Kenyon\footnote{Contact: chris.kenyon@lloydsbanking.com}\ \ and Andrew Green\footnote{Contact: andrew.green2@lloydsbanking.com} }
\title{Efficient XVA Management: Pricing, Hedging, and Allocation using Trade-Level Regression and Global Conditioning\footnote{\bf The views expressed are those of the authors only, no other representation should be attributed.   Not guaranteed fit for any purpose.  Use at your own risk.}}
\date{22 December 2014\vskip5mm Version 1.20}

\maketitle

\begin{abstract}
Banks must manage their trading books, not just value them.  Pricing includes valuation adjustments collectively known as XVA (at least credit, funding, capital and tax), so management must also include XVA.  In trading book management we focus on pricing, hedging, and allocation of prices or hedging costs to desks on an individual trade basis.  We show how to combine three technical elements to radically simplify XVA management, both in terms of the calculations, and the implementation of the calculations.   The three technical elements are: trade-level regression; analytic computation of sensitivities; and global conditioning.  All three are required to obtain the radical efficiency gains and implementation simplification.  Moreover, many of the calculations are inherently parallel and suitable for GPU implementation.  The resulting methodology for XVA management is sufficiently general that we can cover pricing, first- and second-order sensitivities, and exact trade-level allocation of pricing and sensitivities within the same framework.  Managing incremental changes to portfolios exactly is also radically simplified.
\end{abstract}

\section{Introduction}

Banks must calculate, and manage, valuation adjustments across their entire trading portfolio.  Valuation adjustments are collectively known as XVA.  XVA includes the effects of credit (CVA, DVA) \cite{Gregory2009a,Kenyon2013a}, funding (FVA, MVA) \cite{Burgard2013a,Green2014a}, capital (KVA) \cite{Green2014d}, and Tax (TVA) \cite{Kenyon2014c}.  

XVA management includes pricing, hedging, and allocation.  Allocation means the allocation of XVA, and XVA hedging costs to desks.  Hedging costs require the computation of first-order sensitivities such as delta and vega.  Hedging costs may include second-order sensitivities, such as interest rate-credit cross-gamma.  Allocation must be carried out on an incremental basis during daily trading.  Pre-trade these incremental cost allocations are part of the pricing and trading decision.

Here we provide a set of analytically rigorous methods for efficiently managing XVA based on a new analytic approach.  This analytic approach combines three elements: trade-level regression; analytic computation of sensitivities; and global conditioning.  Our combination of these elements enables orders of magnitude improvements on computation times, and orders of magnitude reductions in system implementation times and costs.  Technically this paper generalizes \cite{Green2014a} from MVA to XVA and adds sensitivities and allocation, it also makes explicit elements implicit in \cite{Green2014c}.

Regression-based pricing for CVA was developed in \cite{Cesari2010a,Antonov2011a} based on \cite{Longstaff2001a}. However these methods were not applied for all trades, which is what we do here.  By all trades we mean non-callable trades and European-callable trades, as well as Bermudan- and American-callable trades.  Using regressions as part of sensitivity calculations was introduced in \cite{Wang2009a} but not applied to XVA.  The sensitivities covered in \cite{Wang2009a} were limited to those expressed by the regression variables themselves.  Instead we cover all sensitivities by including sensitivities of the underlyings to hedging instruments via the chain rule of differentiation.  Sensitivities of underlyings to hedging instruments can be calculated using Analytic Derivatives (AD) as in \cite{Broadie1996a,Giles2006a} or by using Adjoint Algorithmic Differentiation (AAD).  AAD was introduced in the CVA context by \cite{Capriotti2011a,Capriotti2014a} based on \cite{Naumann2012a}.   We use A/AD as a label for both methods.

Allocation methods have been developed for capital and CVA, notably in \cite{Tasche2008a,Pykhtin2011a}.  These have used Euler allocation which is based on the properties of homogeneous polynomials, with extensions to deal with collateralized trades.  Trade level allocation was introduced for CVA in \cite{Pykhtin2011a} but we go further to XVA.  This requires a general global conditioning approach to cover MVA when based on Expected Shortfall (ES) or Value-at-Risk (VAR) as well as CVA, DVA, and FVA.    Exact additive allocation of XVA sensitivities is simplified using trade-level regressions, and A/AD on underlyings with global conditioning.  By using regressions for loss given default (LGD) and probabilities of default (PD) we handle the entire XVA computation, and sensitivities simply.  The computations are non-linear (multiplication of regressions for value, LGD, and PD)  but very simple analytically.  This analytic simplicity carries over to sensitivities and exact additive allocation.  Our approach additionally allows exact calculation of incremental changes of XVA, XVA sensitivities, and XVA allocation, with minimal effort.  Such efficient XVA management capabilities are new to the literature.

The contribution of this paper is showing how to combine trade-level regression, analytic computation of sensitivities, and global conditioning to make XVA management computations radically more efficient.  Each element alone is useful but only in combination does the step-change in efficiency of computation and implementation appear.     This combination is highly suitable for parallel implementation on GPUs enabling further speed-ups.

We start with a set of examples of XVA management cases on toy problems.  Once this intuition is built, the subsequent section provides the mathematical generalization, and the final section concludes. 

\section{Examples}

Before we give the general mathematical development we introduce the key elements using examples.  Our objective is to build intuition on how each element works and how they fit together.  Each example is introduced with its objective.

\subsection{Values with Trade-Level Regression}

We demonstrate that computing portfolio prices from the portfolio regression is identical to computing the sum of the individual trade regressions.  The basis function coefficients of the portfolio regression are the sums of the trade basis function coefficients.  This identity holds in general when regressions are linear in the coefficients of the basis functions.

\begin{figure}[htbp]
	\centering
		\includegraphics[width=0.8\textwidth,clip,trim=0 525 150 50]{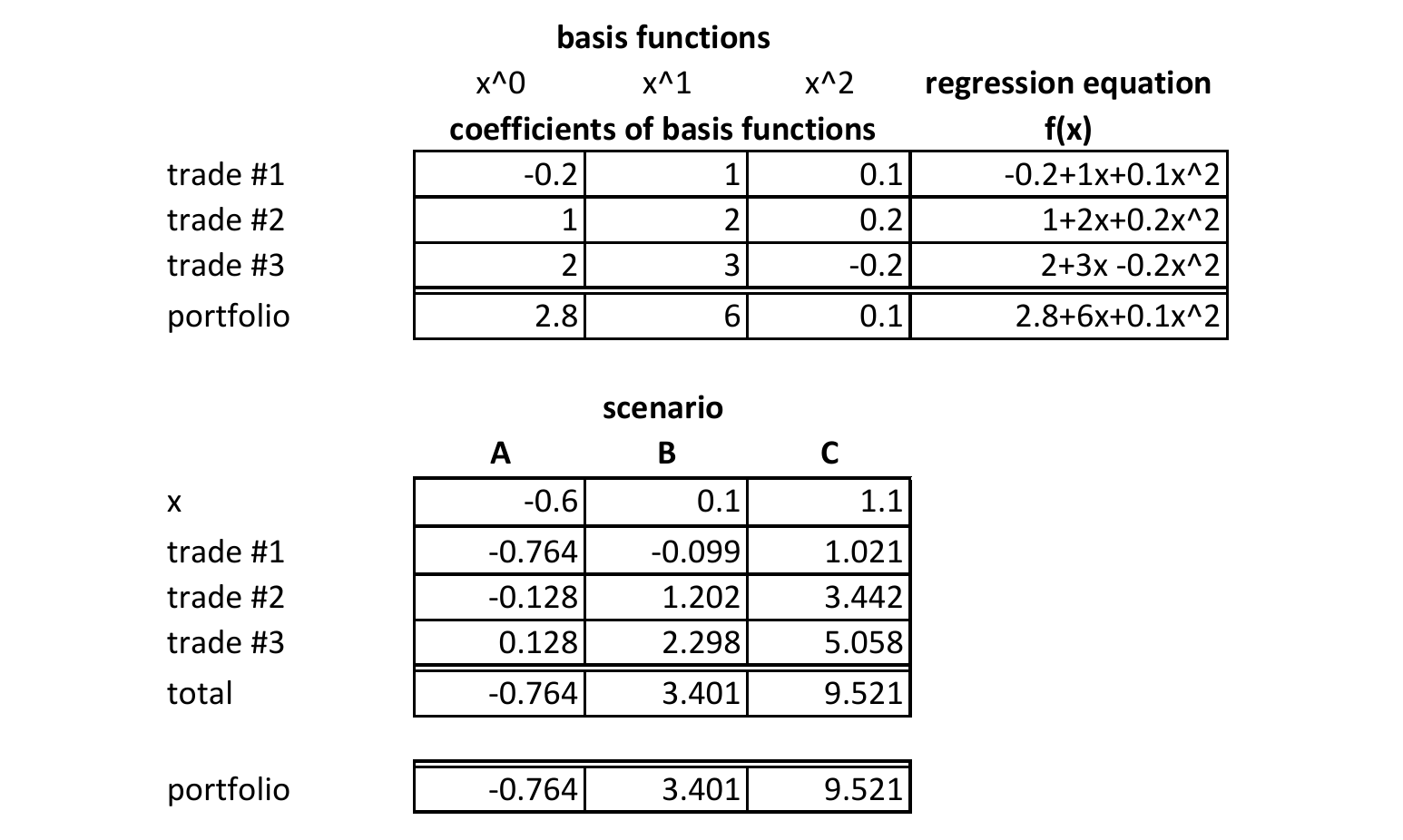}
	\caption{Trade-level regression example.  When regressions are linear in their coefficients calculating from the sum of the basis function coefficients (the portfolio regression) is identical to calculating the sum of each trade regression.  }
	\label{x:regression}
\end{figure}

We have one regression equation for each of the three trades.  Each regression equation has three basis functions $\{x^0,\ x^1,\ x^2\}$, so is quadratic in the underlying $x$, but linear in the coefficients of the basis functions.  Figure \ref{x:regression} considers three scenarios $\{A, B, C\}$.  These might be Monte Carlo realizations and there would generally be many thousands of them.  We see that pricing with individual trade-level regressions, or pricing with the sum of the regression equations is identical in all scenarios.  Each scenario $\{A, B, C\}$ is distinguished by the value of the underlying $x$.  $x$ might be a stock price or the price of an interest rate swap, etc.   

\FloatBarrier

\subsection{First-Order Sensitivities with Trade-Level Regression\label{ss:delta}}

This example has three objectives: 1) show how to compute the first-order sensitivity of each trade with respect to any calibration instrument $s$; 2) demonstrate that the resulting trade-level sensitivity regressions add as before to the portfolio sensitivity regression; 3) demonstrate that the implementation effort is radically reduced w.r.t. non-regression-plus-A/AD approaches.  We reuse the setup from the previous example.

To get the first-order sensitivity of any trade regression $f(x)$ to the calibration instrument $s$ we use the chain rule:
\ben
\frac{\partial f(x)}{\partial s} = \frac{\partial f(x)}{\partial x}  \frac{\partial x}{\partial s}		\label{e:chain}
\een
This is also valid for the portfolio regression.  In addition, the coefficients of the basis functions of the sensitivity regressions add, as before, to give the coefficients of the basis functions of the portfolio regression.  Equation \ref{e:chain} is valid for {\it all} first-order sensitivities.  Each separate first-order sensitivity is distinguished by different $s$ and so different $\frac{\partial x}{\partial s}$.

The derivative of the regression w.r.t. the underlying is generally trivial to compute analytically.  The derivative of the underlying w.r.t. the calibration instrument is generally more involved and can be tackled using A/AD.  

By separating the derivative into two parts we radically reduce implementation effort of A/AD because the second part $\frac{\partial x}{\partial s}$ is the same for all trades, and the first part $\frac{\partial f(x)}{\partial x}$ is generally trivial analytically.  Note that both a regression approach and A/AD are required.

\begin{figure}[htbp]
	\centering
		\includegraphics[width=0.8\textwidth,clip,trim=0 515 150 55]{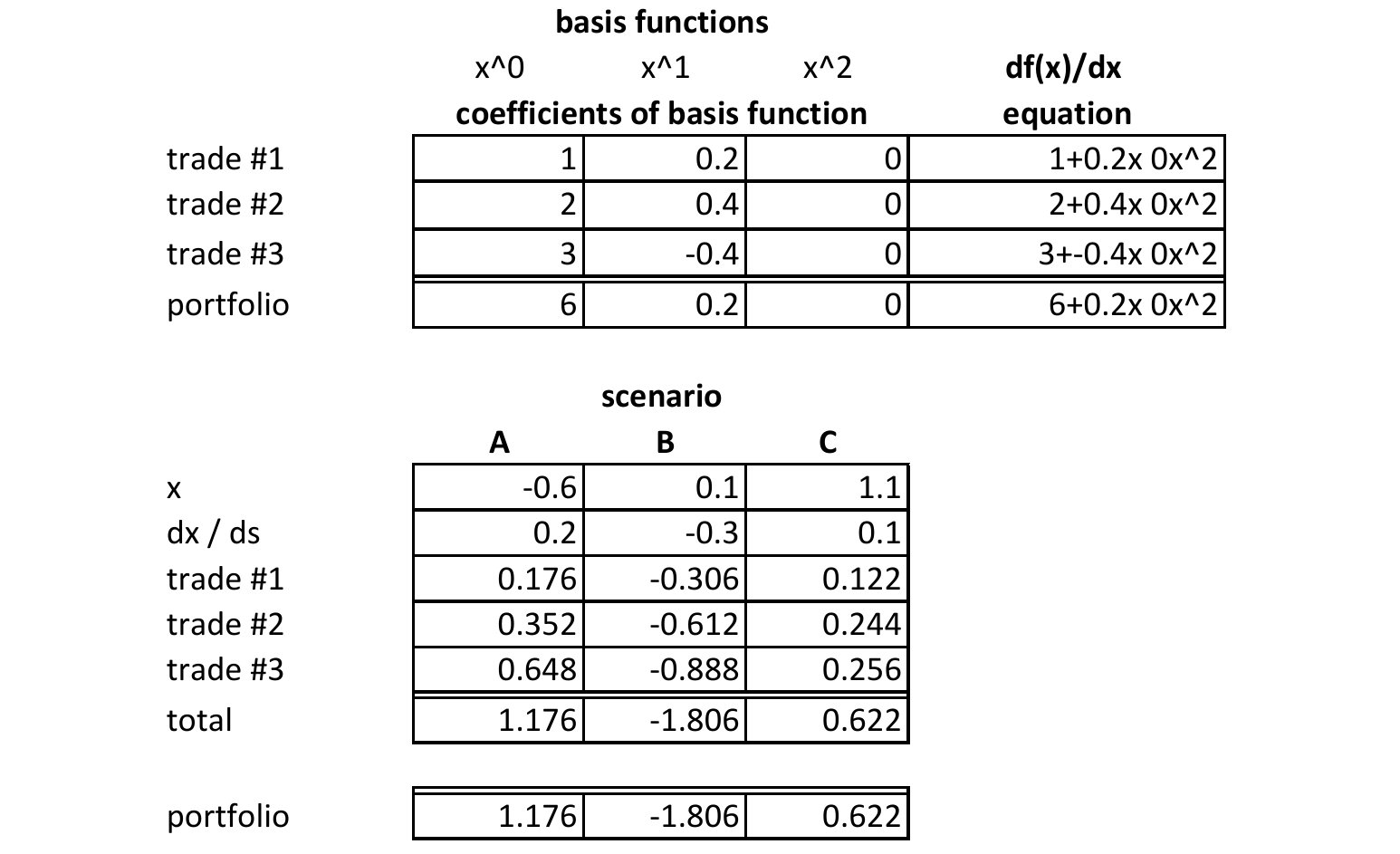}
	\caption{First-order sensitivity example, $s$ is any calibration instrument.  Calculation from trade sensitivity equations or portfolio sensitivity equation is identical.  Portfolio sensitivity equation is identical to the sum of the individual trade equations.}
	\label{x:delta}
\end{figure}

\subsection{Second-Order Sensitivities with Trade-Level Regression}

This example demonstrates that second-order sensitivities work just like first-order sensitivities.  

To get the second-order sensitivities we apply the product rule and the chain rule to Equation \ref{e:chain}, to obtain the derivative with respect to a second calibration instrument $r$:
\ben
\frac{\partial^2 f(x)}{\partial s \partial r} 
= \frac{\partial^2 f(x)}{\partial x^2}  \frac{\partial x}{\partial r} \frac{\partial x}{\partial s}
+ \frac{\partial f(x)}{\partial x}  \frac{\partial^2 x}{\partial s \partial r}
\label{e:gamma}
\een
Equation \ref{e:gamma} is valid for and second-order sensitivity.  Note that $\frac{\partial^2 f(x)}{\partial s \partial r}$ is just a polynomial in $x$ with constant coefficients, just as $\frac{\partial x}{\partial s}$ was.  Hence all the properties and comments for the first-order sensitivities carry over for second-order sensitivities.  This includes the reduction in implementation costs because we will already have $\frac{\partial x}{\partial r}$, and $\frac{\partial^2 x}{\partial s \partial r}$ will be the same for all trade regressions.

\FloatBarrier
\subsection{CVA Calculation and Exact Allocation}

This example introduces global conditioning, and demonstrates its use for CVA and exact allocation.  Global conditioning means that we first identify the scenarios that contribute to a computation using the portfolio regression.  We subsequently calculate using {\it only} those scenarios.  Within the selected scenarios computations are additive because conditional expectation is a linear operator \cite{Shreve2004a}.  

\begin{figure}[htbp]
	\centering
		\includegraphics[width=0.8\textwidth,clip,trim=25 575 75 50]{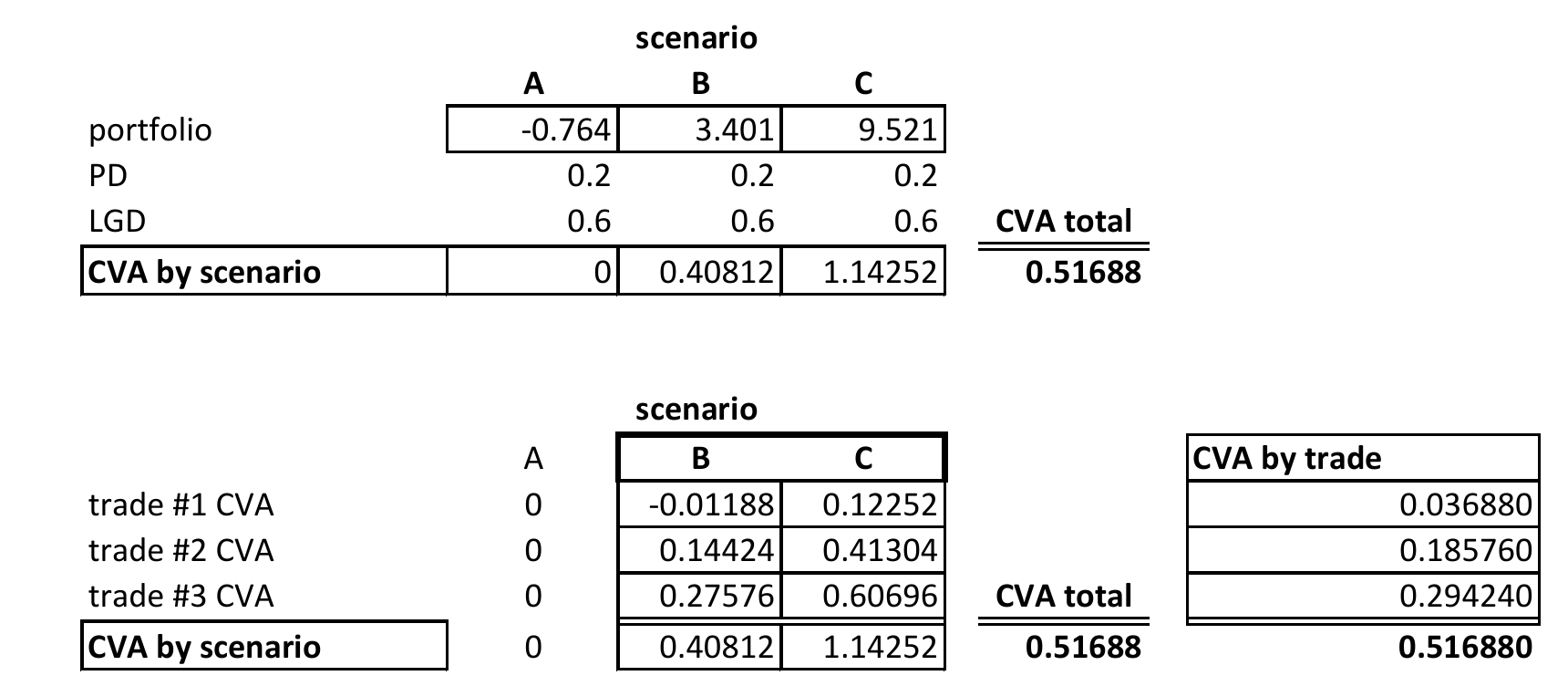}
	\caption{CVA example with exact allocation via global conditioning.  Global conditioning means that we only use scenarios identified as contributing to CVA in the trade-level calculations, i.e. B and C, and set the contributions from scenario A to zero.  So the calculation of CVA by trade using global conditioning provides exact, and additive, allocation.}
	\label{x:cva}
\end{figure}

We use the same portfolio and scenarios as before for this CVA example.  Looking at the portfolio values in Figure \ref{x:regression} we see that only scenarios B and C contribute to CVA because that is where the portfolio has positive value.  Figure \ref{x:cva} shows the CVA calculation: PD is probability of default; LGD, loss given default.  Trade-level contributions to CVA sum exactly. Thus global conditioning provides exact trade-level CVA allocation.  

Re-allocation of trade-level CVA to different desks, or for different reports, is trivial because trade-level contributions sum exactly.

\subsection{Incremental CVA and Exact Allocation}

Suppose we have an incremental trade (trade \#4).  Now for CVA we know that provided its value in scenario A is less than $0.764$ then scenario A will not become relevant for CVA.  We also know that if the value of the new trade is greater than $-3.401$ in scenario B and greater than $-9.521$ in scenario C, then these two scenarios will remain relevant for CVA.  Thus, within these (one-sided) bounds the trade level allocation of CVA will be {\it unchanged} by the new trade.  Its own contribution can be computed independently (conditioned on scenarios B and C).

If the value of the incremental trade is outside the bounds we have identified from the portfolio value in each scenario then we will need to calculate trade values in the newly-relevant scenario.  The previous trade-level values in each previously-relevant scenario will either remain valid, or be set to zero.  Thus re-calculation is essentially trivial, and exact, both for CVA itself and for trade-level allocation.  The same result holds for trade-level contributions to first- and second-order sensitivities.  We say {\it contributions} because sensitivities combine trade parts and parts from default probabilities and recovery rates, as we now show.

\subsection{CVA First-Order Sensitivities and Exact Allocation}

This example shows how global conditioning, regression, and A/AD combine in the computation of first-order sensitivities for CVA.  It also demonstrates exact trade-level allocation of first-order CVA sensitivities.  We will see that the computational and implementation advantages observed previously observed also apply here.

In this example we take LGD as a function of two underlyings $x$ and $y$ and the PD as a function of a single underlying $y$.  Both LGD($x,y$) and PD($x$) are given by regression equations that are linear in the coefficients of their basis functions.  

Since sensitivities are infinitesimal calculations, the scenarios that contribute to CVA sensitivities are exactly those that contribute to CVA value.  Thus the scenarios identified by conditioning on positive portfolio value are still the ones we use for computation of CVA sensitivity.  Within each scenario we have selected we calculate first-order CVA sensitivity with respect to a calibration instrument $s$ as:
\bean
\frac{\partial \{ f(x)\ \lgd(x,y)\ \pd(y)\} }{\partial s} &=& \frac{\partial f(x)}{\partial x} \frac{\partial x}{\partial s}	 \ \lgd(x,y)\ \pd(y)\nonumber	\\
&&{}+  f(x)\left(\frac{\partial \lgd(x,y)}{\partial x} \frac{\partial x}{\partial s} + 	 \frac{\partial \lgd(x,y)}{\partial y} \frac{\partial y}{\partial s}  \right) \pd(y) \nonumber  \\
&&{}+ f(x)\ \lgd(x,y) \frac{\partial\pd(y) }{\partial y} \frac{\partial y}{\partial s}	 \label{e:cva}  
\eean
Although $f(x)\ \lgd(x,y)\ \pd(y)$ is a non-linear expression in $x$ and $y$, it is linear in the coefficients of $x$ and $y$.  Essentially we can  simply regard it as a new function:
\be
 g_*(x,y)=f_*(x)\ \lgd(x,y)\ \pd(y)
\ee
Where $*$ can be ``trade'' or ``portfolio''.  This new function is linear in the coefficients of its basis functions provided that its components $f(x)$, $\lgd(x,y)$, $\pd(y)$ were\footnote{We use polynomials as basis functions and, technically, polynomials over the reals form a commutative ring.  This ring is closed under differentiation.}.  This equation is valid for each trade individually, and for the portfolio.  Thus, within the scenarios chosen, this example is identical to the example in Section \ref{ss:delta}.  

Putting this all together the first order sensitivity of CVA to the calibration instrument $s$ is, in our example:
\bean
\frac{\partial \CVA }{\partial s} &=&\frac{1}{3} \sum_{f_{\text{portfolio}}(x)\ge 0}  \frac{\partial g_{\text{portfolio}}(x,y) }{\partial s}  \label{e:cvaDelta} \\
&=&\frac{1}{3} \sum_{\text{scenario A or B}} \left(\frac{\partial g_{\text{portfolio}}(x,y)}{\partial x} \frac{\partial x}{\partial s} + 	 \frac{\partial g_{\text{portfolio}}(x,y)}{\partial y} \frac{\partial y}{\partial s}  \right) \nonumber
\eean
where we only compute over the two scenarios where the portfolio, $f(x)_{\text{portfolio}}$, is positive.  The factor of one third comes from averaging over all scenarios, although only two provide any contributions.  Note that we retain the separation of trivial differentiation for $g(x,y)$ and the complex derivatives  $\frac{\partial y}{\partial s}$ and $\frac{\partial x}{\partial s}$.  As before the more complex derivatives will require A/AD, but they are only needed for the underlyings ($x,y$).  The derivatives of the underlyings are common for all trades thus we retain the implementation simplification.

We can expand Equation \ref{e:cvaDelta} to see the trade contributions
\bean
\frac{\partial \CVA }{\partial s}  &=&\frac{1}{3}\sum_{f_{\text{portfolio}}(x)\ge 0} \sum_{\text{trades}}\frac{\partial g_{\text{trade}}(x,y) }{\partial s} \nonumber\\
&=&  \sum_{\text{trades}}\frac{1}{3} \sum_{f_{\text{portfolio}}(x)\ge 0}  \frac{\partial g_{\text{trade}}(x,y) }{\partial s} \label{e:cvaDeltaTrade}\\
 &=&\sum_{\text{trades}} \frac{1}{3}\sum_{\text{scenario A or B}} \left(\frac{\partial g_{\text{trade}}(x,y)}{\partial x} \frac{\partial x}{\partial s} + 	 \frac{\partial g_{\text{trade}}(x,y)}{\partial y} \frac{\partial y}{\partial s}  \right) \nonumber
\eean
Note that the global conditioning, on the portfolio value $f_{\text{portfolio}}(x)$, is used and that we can invert the order of summations.  The trade sensitivity contributions are exactly their CVA sensitivity allocations.

Second-order sensitivities for CVA, and their exact allocation, work just like first order sensitivities for CVA.

\FloatBarrier
\subsection{Expected Shortfall Calculation, Sensitivities, \\ and Exact Allocation}

This examples demonstrates that our combination of trade-level regression, A/AD, and global conditioning can be applied to Expected Shortfall (ES).  Expected shortfall is a risk measure based on conditional expectation, the average above a given percentile. Typically ES(97.5\%) will be of interest, so with 2500 scenarios the average will be taken over the 2.5\%\ that show the largest losses, i.e. an average over the 62 worst-loss scenarios.  For reasons of space we consider ES(60\%) with five scenarios, so we average over the losses of the worst two scenarios.

\begin{figure}[htbp]
	\centering
		\includegraphics[width=0.9\textwidth,clip,trim=0 500 50 50]{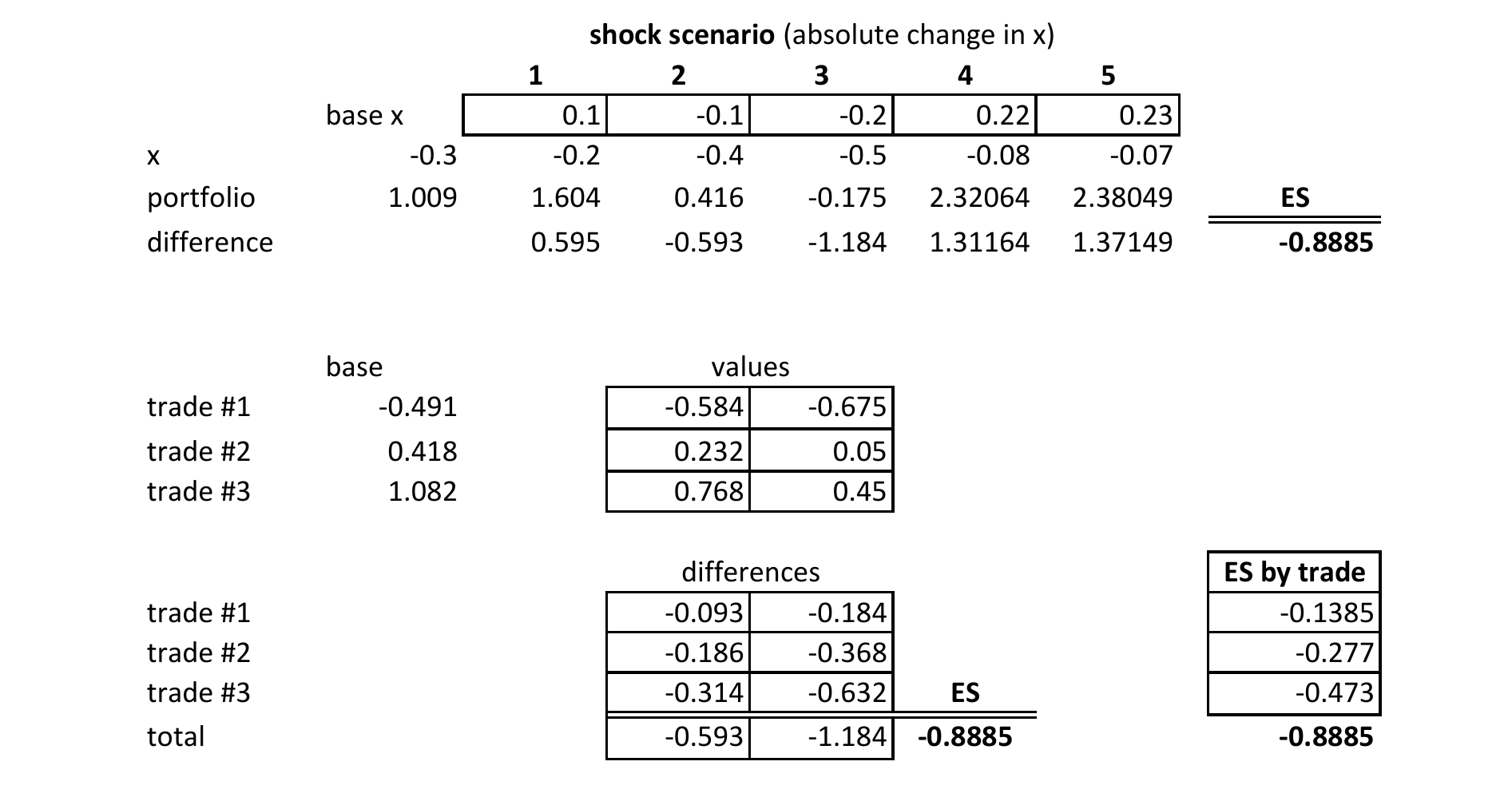}
	\caption{Exact allocation of Expected Shortfall (ES) using global conditioning and regression.  Once we know from portfolio calculation which shock scenarios contribute to ES (shock scenarios 2 and 3 here for ES(60\%) we only need to compute trade values for those scenarios, and we can compute them easily using regression.  Note that although all the trades had losses in both ES-used scenarios, this may not happen in general.}
	\label{x:ES}
\end{figure}

Since ES  is based on conditional expectation we can apply similar logic to its calculation and exact allocation as shown in Figure \ref{x:ES}. 
There are two points to note: firstly the trade-level allocation is exact; secondly we only compute the trade values where they are needed for ES.  In general this will result in speed-ups of roughly 40-fold for trade-level allocation for ES(97.5\%).

The logic applied here for ES is also applicable for VaR, seen as an ES with upper and lower percentile limits. 

First-order and second-order sensitivities for ES can be calculated in the same way as for CVA, i.e. contributions will be additive and exact.  This applies for all orders of sensitivities. 

\subsection{Lifetime Expected Shortfall Calculation, Sensitivities,\\ and Exact Allocation}

The logic that we have applied to CVA and to ES can be combined for fast computation of lifetime ES, its sensitivities, and its exact allocation.  Different shocks may occur in different scenarios when shocks are normalized by the difference in market levels from the original shock generation, or for other reasons \cite{Kenyon2014a2}.  By averaging the trade-level ES-allocation in each of the lifetime scenarios  we obtain trade-level allocation of Lifetime ES.  Since we are averaging we assume that each scenario has appropriately discounted, or otherwise adjusted (e.g. by survival probability), values.

Just as in previous cases we can compute first- and second-order sensitivities of Lifetime ES using trade level regression.  We use the ES-relevant scenarios identified from portfolio ES computation in each lifetime scenario.  Because we are within the global conditioning we, again, have additivity of trade-level sensitivity contributions.  Thus allocation, and re-allocation, are trivial.  Hence, for example, we can provide trade-level allocation of MVA sensitivities when this is based on ES (or VaR) type computations.

%\FloatBarrier
\section{Methodology}

Having given a set of motivating examples we now develop the general theory.  We start by formally introducing the three key technical elements, trade-level regression, analytic sensitivity computation, and global conditioning.  We then combine all three for XVA management.

\paragraph{Trade-Level Regression}
We express every trade across the whole portfolio in terms of a set of basis functions.  That is, we regress the value of each trade across an expanded state space against the basis trade set, at every time point of interest (including $t=0$), which we term stopping dates.  For trades that are Bermudan-callable this is done using Early-Start Longstaff-Schwartz \cite{Longstaff2001a,Wang2009a}, for those that are not Bermudan-callable we can apply the simpler Augmented State Space approach \cite{Green2014a}.  We are interested in many time points for XVA calculations because these involve integrals over time \cite{Burgard2013a,Green2014d}.

For each trade $U_i,\ i=1,\ldots |\Pi|$ in the overall portfolio $\Pi$,  we have:
\bean
U_i(t_k; \xi) &=& \sum_{l=1}^{|f_{*,k}|} a_{i,j,k} f_{l,k}(B_{\bar{j},k}(t_k; \xi)) + \epsilon_{i,k}(t_k; \xi) \qquad \forall\ k=1,\ldots |K| \nonumber\\
&&   \xi \in \Xi(t_k)  \label{e:reg}
\eean
where $f_{l,k}(B_{\bar{j},k})$, are $|f_{*,k}|$ functions of the $|B_{*,k}|$ basis instruments at stopping date $k$, and there are $|K|$ stopping dates from $t=0$ to the last date of interest, say the last cashflow date of the portfolio, $t_k$.  $\epsilon_*$ expresses the regression error at a point $\xi$ within the augmented state space $\Xi(t_k)$ at time $t_k$.  $\bar{j}$ indicates that each $f()$ may depend on an arbitrary subset of the basis instruments.  Apart from standard regularity conditions $f()$ have no restrictions. The augmented state space is created either by Early Start for a simulation \cite{Wang2009a}, or by direct augmentation \cite{Green2014a}.  The $a_*$ are constants.

In Equation \ref{e:reg} it is critical that only the $a_{i,j,k}$ depend on the trade.  The $f_{l,k}$ functions and the $B_{\bar{j},k}$ basis instruments do not depend on the trade.  This means that whilst the basis instruments and function may be arbitrarily complex, they are common for all trades.  This is not a significant restriction because of the finite precision of computation.

With a sufficient number of basis functions $\epsilon_{j,k}(t_k; \xi)$ can be made arbitrarily small within the state space of interest \cite{Longstaff2001a} and we do not include it further.  For a portfolio of swaps out 30 years \cite{Green2014a} showed convergence for a few tens of basis functions for lifetime MVA calculation, but each example will require investigation.  \cite{Gerhold2011a} provides general asymptotic results on convergence.  We have picked a linear regression setup for clarity (linear in $a_*$, but not in $B_*$) but more complex versions are possible.  This is linear because once the points on the state space are known (or chosen by some mechanism) then the $f_*()$ are simply numbers and the $a_*$ can be solved for.

\paragraph{Analytic Sensitivities Computation}
We use  analytic derivatives from regressions, together with analytic or algorithmic derivatives of underlyings, to obtain sensitivities.  In using analytic derivatives of the regressions for sensitivities, as in \cite{Wang2009a}, we depend on the regression being a good representation of the value function.  Convergence of the regression to the value function itself has been extensively studied, both for diffusions \cite{Glasserman2004b} and for L\'evy processes \cite{Gerhold2011a}.  Convergence of the derivatives is covered in Theorem 1 of \cite{Wang2009a}.  Both AD and AAD may be used for derivatives of underlyings with respect to calibration instruments and these techniques have been extensively investigated \cite{Broadie1996a,Giles2006a,Naumann2012a}.

\paragraph{Global Conditioning}
We compute using global conditioning.  By global conditioning we mean that we use global criteria to select scenarios (Monte Carlo, and/or VAR or ES), and then we compute only on those scenarios.  One example of a global criterion is the sign of the value of the portfolio.  We might then use the scenarios this selects to compute trade-level sensitivities.  Technically we are using the linearity property of conditional expectation over filtered probability spaces.   Valuation adjustment are typically additive both {\it within} conditional expectations and {\it across}  times.  This depends only on the properties of conditional expectation, it is independent of the particular scenario or scenarios that may be selected.  

Consider a probability space \probSpace, where $\Omega$ is the universe of events, \F\ \ is a filtration on $\Omega$ and \Pbb\ is a probability measure on \F.  Let $X(t,\omega(t)),Y(t,\omega(t))$ be random variables defined on \probSpace, \G\ be a sub-filtration of \F, where $\omega(t)\in\G(t)$, and   $\alpha(t)$ is a deterministic scalar, then is is an elementary result from Definition 2.3.1 and Theorem 2.3.2 in \cite{Shreve2004a} that:
\ben
\E[X(t,\omega(t)) + \alpha(t) Y(t,\omega(t)) | \G(t)] = \E[X(t,\omega(t)) | \G(t)] + \alpha(t) \E[ Y(t,\omega(t)) | \G(t)] \qquad \forall t
\een
So, given a discount bond price $D(t,\omega(t))$, also defined on \probSpace, it is obvious (with appropriate regularity conditions) that:
\begin{flalign}
\int_{t=0}^T \E[  (X(t,\omega(t)) + \alpha(t) Y(t,\omega(t)))D(t,\omega(t)) | \G(t)] dt \qquad\qquad\qquad\qquad\qquad\qquad&  \nonumber\\
= \int_{t=0}^T\E[X(t,\omega(t))D(t,\omega(t)) | \G(t)] dt \qquad\qquad\qquad\qquad\quad \nonumber\\
{}+  \int_{t=0}^T  \alpha(t) \E[ Y(t,\omega(t))D(t,\omega(t)) | \G(t)] dt \qquad\qquad\qquad\nonumber
\end{flalign}
\G\ may have sub-filtrations within it in turn, that is, \probSpace\ may contain nested probability spaces.  Thus we have demonstrated actually linearity of computation of lifetime costs of conditional quantities, for example, Exposures (including survival weighted), VAR, Expected Shortfall, etc.  This is key for allocation and re-allocation without re-simulation using trade-level regression.  It also means that allocation methods can be exact, given \G.  

One sub-filtration of $\Omega$ will typically be the risk-neutral filtration, another will be its augmentation.  For example at $t=0$ the augmented state space $\Xi(0)$ contains a wide set of events that are used to obtain a regression suitable for use in VAR or ES. $\Xi(t)$  is driven by a set of events with given probabilities (often a set of historical events).

\paragraph{Implementation}  Consider the linear equation:
\be
A x = b
\ee
where $A$ is a non-square data matrix, and $x,b$ are vectors, $b$ data, and $x$ the unknowns.  Now suppose that $A$ contains the values of the basis instruments for a set of scenarios, and $b$ the values of the target instrument over the same set of scenarios.  A least-squares solution for $x$ (the coefficients of the basis instruments) can be obtained using a Singular Value Decomposition (SVD) of $A$ followed by a back-substitution step \cite{Press2007a,Strang2009a}.  

Whilst there will be many instruments in the portfolio, the (overall) set of basis instruments is fixed, so the SVD of $A$ need only be done once (or once per stopping date).  The back-substitution step for every trade in the portfolio at every stopping date (prior to instrument maturity) can be done in parallel, and is an ideal candidate for GPU implementation.  Bermudan-callable instruments need further steps \cite{Longstaff2001a} but still contain regression steps using $A$, so that is common.  Fast GPU implementations of American options have been demonstrated in benchmarking \cite{Denmouth2014a}.

We now apply the technical points to XVA management and show how they simplify, and accelerate, computation.

\subsection{CVA, DVA, FVA: Pricing, Sensitivities, and Allocation}

\begin{table}[tbp]
	\centering
		\begin{tabular}{l|c|c}
		  xVA${}_*^\bigtriangledown$						& $*$ & $\bigtriangledown$ \\ \hline
			CVA 					& $C$ & ${}+{}$   \\
			DVA 					& $B$ & ${}-{}$ \\
			FCA 					& $B$ & ${}+{}$ \\
			FVA = DVA+FCA	& $B$ & (blank) 
		\end{tabular}
	\caption{Alternatives for $*$ and $\bigtriangledown$ that select different xVA possibilities in Equation \ref{e:xva} in the text (equation adapted from Burgard and Kjaer 2013).}
	\label{t:xva}
\end{table}
We start from a generic valuation adjustment of an uncollateralized netting set between a bank, $B$, and its counterparty $C$.  The valuation adjustment can be for CVA, DVA, or FVA so we label it xVA.  Its equation from  {\it Strategy I: Semi-replication with no shortfall at own default} in \cite{Burgard2013a} is:
\ben
\xVA_*^\bigtriangledown = -\lgd_* \int_t^T  \lambda_*(u) D_q(t,u) \E_t\left[V(u)^\bigtriangledown \right] du \label{e:xva}
\een
where ${}_*^\bigtriangledown$ determines which xVA this computes, see Table \ref{t:xva}.  $*$ is either $B$ for the bank, or $C$ for the counterparty. $\bigtriangledown$ can select the positive exposure, the negative exposure, or do nothing depending on the particular valuation adjustment. $D_q(t,u)$ is the discount factor between $u$ and $t$ for the rate $q$,  $q=r+\lambda_B+\lambda_C$: $r$ is the riskless rate, and $\lambda_*$ the hazard rate of the bank or counterparty.  $V$ is the unadjusted value of the netting set. 

Let the netting set $V$ be made up of a set of trades $\pi_i$, so Equation \ref{e:xva} becomes:
\ben
\xVA_*^\bigtriangledown = -\lgd_* \int_t^T \lambda_*(u) D_q(t,u) \E_t\left[\left(\sum_i \pi_i(u)\right)^\bigtriangledown \right] du \label{e:xva2}
\een
Now suppose that the the expectation and time integral are both computed using sets of observations (e.g. generated by simulation) we have (using the simplest possible time-integration scheme):
\be
\xVA_*^\bigtriangledown  = - \lgd_*  \sum_{k=1}^{n_k} (t_k-t_{k-1}) \lambda_*(u_k) D_q(t,u_k)  \frac{1}{n_j} \sum_{j=1}^{n_{j}} 
     \left(\sum_i \pi_i(u_k; \omega_{j,k})\right)^\bigtriangledown       
\ee
Where $n_k$ is the number of time steps and $n_j$ is the number of scenarios at each time stpe. $\omega_{j,k}$ represents the realization of the random factors at time $t_k$ in scenario $j$. $\omega_{j,k} \in \G(t_k)$, where $\G$ is the filtration.

Applying Equation \ref{e:reg}, i.e. using regressions for the trade values we obtain:
\bea
\xVA_*^\bigtriangledown  &=& - \lgd_*  \sum_{k=1}^{n_k} (t_k-t_{k-1}) \lambda_*(u_k) D_q(t,u_k)  \\
&& \quad \times \frac{1}{n_j}\sum_{j=1}^{n_{j}} 
     \left(      \sum_i \sum_{l=1}^{|f_{*,k}|} a_{i,l,k} f_{l,k}(B_{\bar{j},k}(t_k; \omega_{j,k}))       \right)^\bigtriangledown   \\
&=& - \lgd_*  \sum_{k=1}^{n_k} (t_k-t_{k-1}) \lambda_*(u_k) D_q(t,u_k)  \\
&& \quad \times \frac{1}{n_j} \sum_{j=1}^{n_{j}} 
     \left(       \sum_{l=1}^{|f_{*,k}|} a_{l,k} f_{l,k}(B_{\bar{j},k}(t_k; \omega_{j,k}))       \right)^\bigtriangledown   \\
a_{l,k} &=& \sum_i a_{i,l,k}
\eea
Since the basis instruments are common for all trades we now have an equation involving only these instruments.  We can now expand the equation for $\xVA_*^\bigtriangledown$ above to remove the non-linearity of the $()^\bigtriangledown$ bracket as:
\bean
\xVA_*^\bigtriangledown &=& - \lgd_*  \sum_{k=1}^{n_k} (t_k-t_{k-1}) \lambda_*(u_k) D_q(t,u_k)  \nonumber \\
&& \quad \times \frac{1}{n_j} \boxed{ \sum_{j=1}^{n_{j}} \I{ {V_{j,k}^\bigtriangledown} } }
            \sum_{l=1}^{|f_{*,k}|} a_{l,k} f_{l,k}(B_{\bar{j},k}(t_k; \omega_{j,k}))  \\
								&=& - \lgd_*  \sum_{k=1}^{n_k} (t_k-t_{k-1}) \lambda_*(u_k) D_q(t,u_k)  \nonumber \\
&& \quad \times \frac{1}{n_j} \boxed{ \sum_{j |V_{j,k}^\bigtriangledown}^{\vphantom{n_j}}  } 
            \sum_{l=1}^{|f_{*,k}|} a_{l,k} f_{l,k}(B_{\bar{j},k}(t_k; \omega_{j,k}))  \label{e:select}
\eean
The boxes indicate the key point of the development --- we compute with respect to a set of global scenarios selected by our conditioning criteria.  This is an example of global conditioning.  We use $\I{V_{j,k}^\bigtriangledown}$ as the indicator function on the sign of the unadjusted future netting set value (see Table \ref{t:xva} for choices).   In Equation \ref{e:select} we have selected, at each time point, those scenarios such that the netting set value satisfies the criteria (i.e. the same selection as the indicator function).  These scenarios will be different for each time point.
  
At this point it may be argued that we have done more work, rather than less.  We have had to value all the original trades under all (augmented) scenarios and at all time points to obtain the regression coefficients.  In addition we have had to calculate the regressions.  

However, xVA is only a first step, and also our target is XVA {\it management}.   We now demonstrate what we have achieved for sensitivities, allocation, and allocation of xVA sensitivities.  In general all of these calculations are more costly than the initial xVA computation.   In the next section we will use exactly the same regressions for MVA computation, sensitivities, allocation and allocation of MVA sensitivities.  Thus even for XVA {\it computation} we will demonstrate significant advantages.

\paragraph{Sensitivities} By sensitivity we mean sensitivity with respect to hedging, i.e. calibration, instruments.  We assume that the two sets are identical.  We further assume that sensitivities are being computed analytically (or algorithmically, we make no distinction on method except that it is not bumping).

Obviously we have immediately reduced the implementation cost of analytic derivatives from all the trade types found in the entire portfolio to the set of basis instruments.  This will usually represent a major saving in implementation time.

The next key observation is that since analytic sensitivities are based on infinitesimal changes, and we compute at finite precision,  the set of scenarios we calculate over, $j | {V_{j,k}^\bigtriangledown}$, is unchanged.  Hence, for a calibration instrument $s$ that we want a sensitivity for:
\be
\frac{\partial \xVA_*^\bigtriangledown }{\partial s} = \ldots \frac{\partial f_{l,k}(B_{\bar{j},k}(t_k; \omega_{j,k})) }{\partial B_{\bar{j},k}(t_k; \omega_{j,k})}
\frac{\partial B_{\bar{j},k}(t_k; \omega_{j,k}) }{ \partial s}
\ee
Alternatively
\be
J_{\xVA_*^\bigtriangledown , s} = \ldots J_{ f_{l,k}(), B_{\bar{j},k}()  }  J_{B_{\bar{j},k}(), s  }
\ee
where $J_{*,*}$ are the Jacobians.  

We have pre-selected the scenarios $j | {V_{j,k}^\bigtriangledown}$ to calculate over so everything is linear --- and since differentiation is a linear operator this remains.  Generally $J_{ f_{l,k}(), B_{\bar{j},k}()  } $ will be analytic because the $f_{l,k}()$ will have been selected for that property.  Typical sets of orthogonal basis functions such as sine and cosine, or Chebeshev polynomials have simple analytic derivatives \cite{Press2007a}.

\paragraph{Trade-Level Allocation of xVA}

Allocation of valuation adjustment prices to desks is a core activity of XVA desks. Trade-level allocation of xVA is given directly from Equation \ref{e:select} by considering the contribution of each trade in terms of its regression coefficients.  For example for trade $i$:
\bea
\xVA_*^\bigtriangledown &=&- \lgd_*  \sum_{k=1}^{n_k} (t_k-t_{k-1}) \lambda_*(u_k) D_q(t,u_k)  \nonumber \\
&& \quad \times \frac{1}{n_j} { \sum_{j |V_{j,k}^\bigtriangledown}^{\vphantom{n_j}}  } 
            \sum_{l=1}^{|f_{*,k}|} a_{i,l,k} f_{l,k}(B_{\bar{j},k}(t_k; \omega_{j,k}))  \label{e:cvaA}
\eea
Where the difference from Equation \ref{e:select} is that we now use the trade $i$'s regression coefficients $a_{i,l,k}$ rather than the netting set regression coefficients $a_{l,k}$.  Allocation is both exact and additive using trade-level regression with global conditioning, i.e. computing within the selected scenarios $j | {V_{j,k}^\bigtriangledown}$.  Thus re-allocation, i.e. re-allocation of different trades' xVA to different groupings, e.g. for reporting, is trivial.

\paragraph{Trade-Level Allocation of Sensitivities}
Hedging costs are often derived from sensitivities, thus trade-level allocation of these sensitivities is a core activity of XVA desks.  Since differentiation is a linear operator we can combine the arguments of the previous two sections to observe that  using our regression and conditioning approach, with respect to a calibration instrument $s$:
\be
\frac{\partial \CVA(\Pi^C,t)}{\partial s} = \sum_i \frac{\partial \CVA(\pi_i^C,t)}{\partial s}
\ee
 where we only compute ``trade'' sensitivities {\it within selected scenarios}.  We put trade in quotes because we can arbitrarily create new trade grouping using the additivity of their regression coefficients.  We can thus allocation, and re-allocation, hedging costs freely.  That is, the costs are linear and we only re-allocate sums of scalar numbers to different pots (desks, groups, etc).  In addition we only compute sensitivities at the coarsest level required using the appropriate regression coefficients.

\paragraph{Incremental xVA}  During a trading day there will be continual changes to portfolios and the XVA desk must provide prices for these changes to other desks.  Portfolio changes can be expected to change the conditioning set $j | {V_{j,k}^\bigtriangledown}$.  For xVA the conditioning set is specific to each counterparty.  First note that we have already calculated the unconditioned portfolio values in each scenario:
\be
V_{j,k}^{\text{Unconditioned}}(\Pi) = V_{j,k}
\ee
We follow the same procedure for the new trades as for the existing portfolio by calculating their trade-level regressions.  Now we calculate their values for the same set of overall scenarios as the existing portfolio and calculate the updated conditioning scenarios
\bea
j | V_{j,k}^\bigtriangledown(\Pi(\text{original}) + \Pi(\text{changes})) &=& j|\{ j | V_{j,k}(\Pi(\text{original})) +  j | V_{j,k}(\Pi(\text{changes})) \}^\bigtriangledown  \\
&=& j | V_{j,k}^\bigtriangledown(\Pi(\text{updated})
\eea
To compute the first line above we only need the scenario values of the original portfolio and the changes to the portfolio.  No re-computation of the original portfolio is required.  Thus we can re-compute the CVA without recomputing the original portfolio, we just include the previous values for the additional scenarios.  

Again it appears that we have computed the regression of the changes to the portfolio as extra work.  However, this extra work makes the other XVA elements, and their management (sensitivities and allocation) orders of magnitude faster.

For incremental sensitivities the arguments of the previous sections apply directly.  This is also true for incremental trade-level allocation, and for incremental trade-level allocation of sensitivities.

\subsection{MVA Pricing, Sensitivities, and Allocation}

Central counterparties often require posting of initial margin (IM) which can depend on portfolio VAR or ES \cite{Gregory2014a}.   The lifetime costs of funding this IM is termed Margin Valuation Adjustment (MVA).

As indicated in the Examples section our technique also applies to the lifetime cost of funding initial margin, i.e. margin valuation adjustment (MVA).   Initial margin for trades with central counterparties is often based on VAR and/or ES, so we consider these next.  As shown in the examples, the use of global scenario selection makes VAR and ES {\it computation} additive.  

ES is a conditional expectation by definition.  Thus the derivation from CVA above applies exactly. Furthermore since we have already calculated the trade-level regression functions, and analytic derivatives, for CVA there is no need to re-compute them for ES.  The only difference between CVA and ES is a different condition:
\be
j | \Pbb_{ES}(V_{j,k})\le\alpha
\ee
instead of 
\be
j | {V_{j,k}^\bigtriangledown}
\ee
Once the scenarios are identified the same computations apply.  We have used $\Pbb_{ES}$ for the distribution of portfolio values as used for ES.  This distribution will usually be on sub-filtration of $\Omega$ equivalent to that used for the risk-neutral measure (but obviously with a different measure).  We use $\alpha$ for the percentile of interest, typically 97.5\%\ or similar.

We approach VAR as a limit of ES definitions, i.e.
\be
j | \Pbb_{ES}(V_{j,k})=\alpha  = \lim_{\beta\rightarrow \alpha} \left\{ j | \beta \le \Pbb_{ES}(V_{j,k})\le\alpha \right\}
\ee
Thus the development above for CVA, and ES, also applies to VAR.  In practice, using a simulation, VAR may be calculated from a very small number of scenarios (possibly just one).

Since we have only changed the conditioning scenarios  the arguments from the previous section apply exactly to all the management cases including sensitivity computation and allocation.

\section{Conclusions}

We have shown how XVA management is radically more efficient using a combination of three technical elements: trade-level regression; analytic derivatives; and global conditioning.  All three elements are required for this radical gain in both computational efficiency and implementation efficiency.  By XVA we mean the lifetime costs of CVA, FVA, DVA, and MVA.  The use of regression for KVA is covered in detail in \cite{Green2014c}.  Under XVA management we include pricing, hedging, and allocation of prices and hedging costs.  

Implementation costs are drastically reduced because implementation of analytic derivatives are now only required for the regressions themselves which are trivial to compute --- not for all the trade types in the original portfolio.  AD, or AAD, is only required for the underlyings themselves, a small subset of total trade types.  Due to the inherent parallelism of many parts of this approach it is ideally suited to GPU implementation.

The main limitation of the technique in this paper is that it is essentially a first-order approach in that it does not deal with changes in option exercise boundaries because of XVA interactions.  This is a topic of further research \cite{Green2015a}.

One limitation may appear to be re-computation of regressions during a trading day.  However, we do not expect this to be a limitation because regression functions are valid over the whole state space so should be robust against intra-day changes or market conditions.

This approach makes computation of trade values cheap and fast.  It also separates the computation of the trade prices from the computation of the values of the underlyings.  This means that the dynamics of the underlyings can be almost arbitrarily complex.  Hence a potential limitation on global pricing is removed.  That is, the XVA dynamics can be made as detailed as required to match time-zero pricing.  Thus only a single system is required for both real-time trade pricing and XVA.

\bibliographystyle{chicago}
\bibliography{kenyon_general}

\end{document}